\documentclass[aps,reprint,pra]{revtex4-1}
\pdfoutput=1 
            
\usepackage[utf8]{inputenc}
\usepackage[top=1.0in, bottom=1.5in, left=1in, right=1in]{geometry}

\usepackage{natbib}
\usepackage{graphicx}
\usepackage{verbatim}
\usepackage{amssymb}
\usepackage{amsmath}
\usepackage{amsfonts}
\usepackage{braket}
\usepackage{dsfont}
\usepackage[compat=1.1.0]{tikz-feynman}
\usepackage[english]{babel}
\usepackage{hyperref}
\hypersetup{urlcolor=red, colorlinks=true}  
\usepackage{pgfplots}
\usepackage{url}
\pgfplotsset{compat=1.12}
\usepackage{setspace}
\usepackage{slashed}

\begin{document}

\title{Finite temperature geometric properties of the Kitaev honeycomb model}
\author{ Francesco Bascone$^{1,5,6}$, Luca Leonforte$^{1}$,  Davide Valenti$^{1,2}$, Bernardo Spagnolo$^{1,3,4}$ and Angelo Carollo$^{1,3}$}

\affiliation{$^1$Dipartimento di Fisica e Chimica, Group of
Interdisciplinary Theoretical Physics and CNISM, Universit\`{a} di
Palermo, Viale delle Scienze, Edificio 18, I-90128
Palermo, Italy\\
$^2$IBIM-CNR Istituto di Biomedicina ed Immunologia Molecolare
``Alberto Monroy", Via Ugo La Malfa 153, I-90146 Palermo, Italy\\
$^3$Radiophysics Department, Lobachevsky State University of Nizhni Novgorod, 23 Gagarin Avenue, Nizhni Novgorod 603950, Russia\\
$^4$Istituto Nazionale di Fisica Nucleare, Sezione di Catania, Via
S. Sofia 64, I-90123 Catania, Italy \\
$^5$Dipartimento di Fisica "E. Pancini", Universit\`{a} di Napoli Federico II, Complesso Universitario di Monte S. Angelo Edificio 6, via
Cintia, 80126 Napoli, Italy \\
$^6$Istituto Nazionale di Fisica Nucleare, Sezione di Napoli, Complesso Universitario di Monte S. Angelo Edificio 6, via
Cintia, 80126 Napoli, Italy }

\begin{abstract}
We study finite temperature topological phase transitions of the Kitaev's spin honeycomb model in the vortex-free sector with the use of the recently introduced mean Uhlmann curvature. We employ an appropriate Fermionisation procedure to study the system as a two-band p-wave superconductor described by a BdG Hamiltonian. This allows us to study relevant quantities such as Berry and mean Uhlmann curvatures in a simple setting. More specifically, we consider the spin honeycomb in the presence of an external magnetic field breaking time reversal symmetry. The introduction of such an external perturbation opens up a gap in the phase of the system characterised by non-Abelian statistics. The resulting model belong to a symmetry protected class, so that the Uhlmann number can be analysed. We first consider the Berry curvature on a particular evolution line over the phase diagram. The mean Uhlmann curvature and the Uhlmann number are then analysed, by assuming a thermal state. The mean Uhlmann curvature describes a cross-over transition as temperature rises. In the trivial phase, a non-monotonic dependence of the Uhlmann number, as temperature increases, is reported and explained.
\end{abstract}


\maketitle

\section{Introduction}
\label{intro}

Topological phase transitions (TPTs) have emerged as a major new paradigm, which eludes the ordinary Laundau classification and where phases are characterised by local order parameters and symmetry breaking occurring across criticalities. Topological phases indeed are identified by integer-valued invariants that are constructed out of ground states properties~\cite{Altland1997,Schnyder2008,Ryu2010,Chiu2016}. Topological systems have attracted a great deal of interest on account of their peculiar properties, ranging from topologically protected edge states~\cite{Hatsugai1993}, to quantised current~\cite{Klitzing1980,Thouless1982,Thouless1983,Niu1984,Nakajima2016,Tsui1982}, and excitations with exotic statistics~\cite{Laughlin1983,Arovas1984,Nayak2008}.  There is already a vast literature concerning zero temperature TPTs, where the systems are described by pure states, but few studies have been done in the direction of a consistent mixed states generalisation. Some recent results have shown that it may be possible to characterise topological phases for thermal states~\cite{Avron2011,Bardyn2013,Huang2014,Viyuela2014,Viyuela2014a,Budich2015a,Linzner2016,Mera2017,Grusdt2017,Bardyn2018}. Among these, particularly promising approaches are based on Uhlmann holonomies~\cite{Mera2017,Huang2014,Viyuela2014a,Viyuela2014}, which are a formal generalisation of the Berry phase for mixed states~\cite{Uhlmann1986,Uhlmann1991}. The latter quantity is in fact one of the main ingredients of the topological phases in the pure states case.
In recent works~\cite{Carollo2018,Carollo2018a,Leonforte2019} it was shown that a physical quantity related to the Uhlmann connection, called mean Uhlmann curvature (MUC), is able to provide interesting features about TPT in the mixed state case, accounting for the effect of temperature at thermal equilibrium or for out-of-equilibrium conditions~\cite{Magazzu2015,Spagnolo2018,Valenti2018,Spagnolo2018a,Guarcello2015,Spagnolo2016,Spagnolo2015,Spagnolo2012}. It was also shown that, in 2D symmetry-protected topological systems, it is possible to define a \emph{Uhlmann number}, which is a direct generalisation of the Chern number, used as a topological invariant describing different topological phases at zero temperature~\cite{Leonforte2019,He2018}. However, this so-defined Uhlmann number is only formally analogue to the Chern number, since it is not a topological invariant and it can be non-integer.

Recently, much effort has been dedicated to the study of fault-tolerant quantum computation via topology \cite{Nayak2008,Preskill1997,Kitaev2003,Kitaev2009}. In this context the Kitaev honeycomb model \cite{Kitaev2006}, extensively studied only at zero temperature, shows a rich phase structure that allows both Abelian and non-Abelian anyonic excitations. Non-Abelian anyons are in fact a crucial building block of topological quantum computing, whereby quantum computation is performed by braiding of excitations. The main purpose of this work is to study the Kitaev's honeycomb model at finite temperature using the mean Uhlmann curvature as a main tool. The analysis of finite temperature phase transitions is in fact especially important in the quantum computing framework since this would allow one to understand how the topological concepts can be used at finite temperature, allowing for better practical opportunities. The honeycomb model under consideration shows a phase diagram containing gapped and gapless phases. At first, we introduce an external magnetic field breaking time-reversal symmetry. By this way the system belongs to the symmetry-protected class $D$, which is characterised by a $+$ charge conjugation type symmetry and by the absence of time-reversal and chiral symmetries \cite{Chiu2016}. In this context, one can analyse the system through the Uhlmann number, since the Chern number is the proper zero-temperature topological invariant of such a class. Furthermore, such an external perturbation allows for the existence of non-Abelian excitations and opens a gap in otherwise gapless phase. 
One of the main results of this paper is the analysis of the Uhlmann number behaviour in the trivial phase for small values of the temperature close to the critical point. We find a non-monotonic behaviour, noted earlier in~\cite{Leonforte2019},  which seems a general feature of the class, which can be in principle observed experimentally.
We also study the Berry curvature of the model, both numerically and analytically, in the absence of external magnetic field interactions as a limit when the external coupling tends to zero. This is necessary because in the vanishing external field case the Berry curvature is zero and it is therefore necessary to extend the parameter space. We analyse the Berry curvature only in this case, because the model becomes topologically intrinsic and the Uhlmann number is no longer the quantity of interest.

The paper is organised as follows. In section~\ref{honey} we discuss the spin honeycomb model and its phase diagram. We employ the fermionisation procedure introduced in~\cite{Kells2009}, which has the advantage to give a closed form of the ground state in a BCS form. With this technique the system can be considered as a two-band p-wave topological superconductor and this allows for more convenient calculations and better understanding of the results. 
In section~\ref{berry} we carry out the calculation of the Berry curvature for the ground state, which is unique in the planar geometry, both in the presence and in the absence of an external magnetic field. In section~\ref{meanuhlmann} we calculate the mean Uhlmann curvature and the Uhlmann number to obtain a description of the system at finite temperature, generalising the results for the Berry curvature and for the Chern number in the presence of an external magnetic field acting on the honeycomb lattice. The section~\ref{conclusions} contains the concluding remarks.

\section{Honeycomb model}
\label{honey}

\begin{figure}[h!]
\centering
\includegraphics[width=\linewidth]{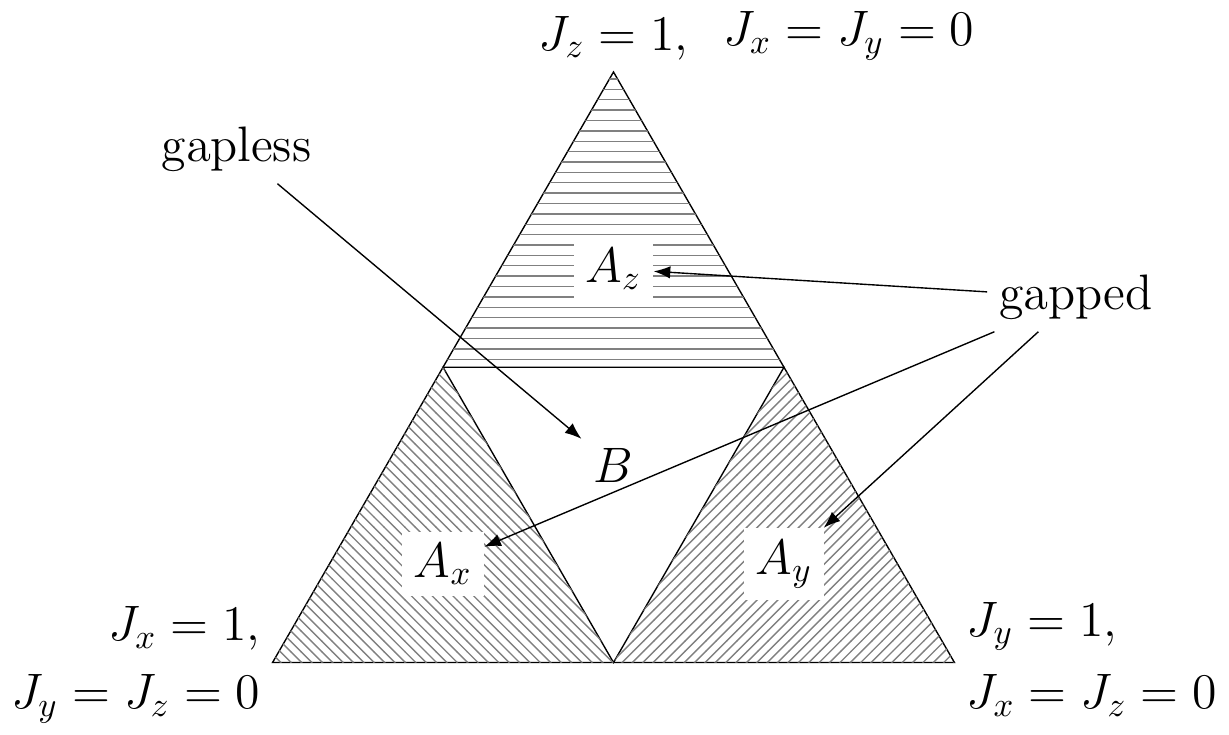}
\caption{Phase diagram of the honeycomb model: the triangle is the section of the positive octant by the plane $J_x+J_y+J_z=1$.}
\label{fig:phasediagram}
\end{figure}

We will consider the Kitaev honeycomb model \cite{Kitaev2006}, which comprises spin-$1/2$ particles arranged on the vertices of a honeycomb lattice. This model can support a rich variety of topological behaviours, depending on the values of its couplings.\\
The Hamiltonian of the system can be written as follows
\begin{equation}
\label{ham}
H=-\sum_{\alpha \in \{x,y,z \}} \sum_{i,j}J_{\alpha}K^{\alpha}_{ij},
\end{equation}
\begin{figure}[t]
\centering
  \includegraphics[width=0.9\linewidth]{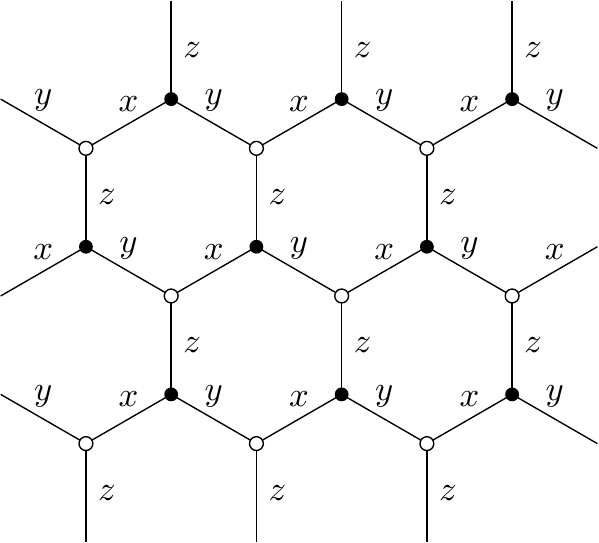}
  \caption{Honeycomb lattice and link-types.}
  \label{fig:honeycomb}
\end{figure}%
\begin{figure}[t]
  \centering
  \includegraphics[width=0.4\linewidth]{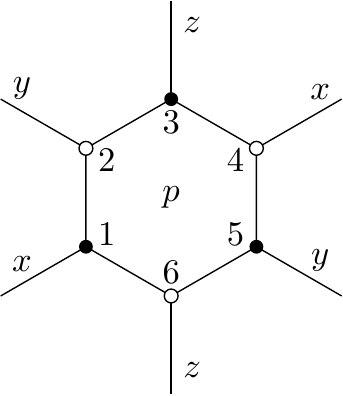}
  \caption{Plaquette structure.}
  \label{fig:plaquette}
\end{figure}
with $K^{\alpha}_{ij}=\sigma_i^{\alpha} \sigma_j^{\alpha}$ denoting directional spin interaction between $i$, $j$ sites connected by $\alpha$-link (see Fig. \ref{fig:honeycomb}), $J_{\alpha}$ are the dimensionless coupling coefficients of the two-body interaction and the $\sigma_i^{\alpha}$ are the Pauli operators. \\
Products of $K$ operators can be used to construct loops on the lattice $K^{\alpha_1}_{i_1, i_2}K^{\alpha_2}_{i_2, i_3} \cdots K^{\alpha_n}_{i_n, i_1}$, and any loop constructed in this way commutes with all other loops and with the Hamiltonian. Therefore, these are good quantum numbers which provide a decomposition of the Hilbert space into direct sum of invariant subspaces. In particular, the shortest loop symmetries are the plaquette operators
\begin{equation}
W_p=K_{12}K_{23}K_{34}K_{45}K_{56}K_{61}=\sigma_1^x \sigma_2^y \sigma_3^z \sigma_4^x \sigma_5^y \sigma_6^z,
\end{equation}
where $p \in \left\{1,2, \dotsc, m\right\}$ is a plaquette index, and $m$ is the number of plaquettes. These $W_p$ operators represent loops around single hexagons and one way to visualise them is to look at the external link-type that is connected to the vertices, e.g. in $\sigma_1^x$, for the external link-type $x$ connected with the vertex 1 (Fig. \ref{fig:plaquette}).\\
The $W_{p}$ are a set of integral of motion whose eigenvalues $\{\pm 1\}$ indentify different \emph{sectors} of the Hilbert space. Each plaquette with $w_p=-1$ is said to carry a vortex, in analogy with the Ising $\mathbb{Z}_2$ gauge lattice theory. Therefore, each sector corresponds to a particular choice of the string of eigenvalues over all the plaquettes $\left\{w_p\right\}|_{p \in \{1,2, \dotsc, m \}}$.\\
In this way, the Hamiltonian can be decomposed as a direct sum over all the configurations:
\begin{equation}
\mathcal{H}=\bigoplus_{\left\{w_p \right\}}\mathcal{H}_{\left\{w_p\right\}}.
\end{equation}
Thus, to solve the problem one needs to find the eigenvalues of the Hamiltonian restricted to a particular sector. There are several ways to exactly solve this problem. According to the Kitaev' s approach, the next step would be to map the spin degrees of freedom to the Majorana fermions and this will require each spin degrees of freedom to be embedded in an extended Hilbert space of dimension four and then to obtain physical states as projections from the eigenstates of the extended Hamiltonian. However, in general this is  quite a daunting task. In some cases, however, it can be more convenient to explore an alternative route, first developed by~\cite{Kells2009,Kells2008}. The latter consists of a Jordan-Wigner (JW) Fermionisation procedure, mapping ``hard-core'' bosons operators to Fermionic operators through string operators. This procedure allows for an explicit construction of the eigenstates of the system.\\
A theorem by Lieb~\cite{Lieb1994} shows that the ground state of the system must lie in the vortex-free sector. By focussing on the vortex-free sector, in a planar lattice geometry, one can exploit the translational symmetry, and use the Fourier transform to derive the energy spectrum. The aforementioned JW transformation results in the following Bogoulibov-deGennes (BdG)-like Hamiltonian,
\begin{equation}
\label{bdghamiltonian}
H=\frac{1}{2}\sum_{\textbf{q}}\left(C^{\dagger}_{\textbf{q}}, \, C_{-\textbf{q}} \right)
 H_{\textbf{q}} \begin{pmatrix}
C_{\textbf{q}} \\
C^{\dagger}_{-\textbf{q}}
\end{pmatrix},
\end{equation}
where,
\begin{equation}
\label{22ham}
H_{\textbf{q}} \equiv \begin{pmatrix}
\xi_{\textbf{q}} & \Delta_{\textbf{q}} \\
\Delta^*_{\textbf{q}} & -\xi_{\textbf{q}}
\end{pmatrix},
\end{equation}
with
\begin{equation}
\begin{aligned}
{} & \xi_{\textbf{q}}=2J_{x}\cos q_x+2J_{y}\cos q_y+2J_z, \\ &
\Delta_{\textbf{q}}=i\beta_{\textbf{q}}=2 i J_{x} \sin q_x+2 i J_{y} \sin q_y.
\end{aligned}
\end{equation}
Here we deal with a Cartesian basis where $\textbf{q} \equiv \left(q_x, q_y \right)$. \\
Thus, the Kitaev honeycomb model is mapped into a spinless fermionic BdG Hamiltonian. The Hamiltonians $H_{\textbf{q}}$ can then be diagonalised via Bogoliubov rotation of the mode operators: $b_{\textbf{q}}=u_{\textbf{q}}C_{\textbf{q}}-v_{\textbf{q}}C^{\dagger}_{-\textbf{q}}$, with
\begin{eqnarray}
u_{\textbf{q}}= \sqrt{\frac{1}{2}+\frac{\xi_{\textbf{q}}}{2\epsilon_{\textbf{q}}}}=\sqrt{1+\frac{J_z}{\epsilon_{\textbf{q}}}}, \\
v_{\textbf{q}}=-i\sqrt{\frac{1}{2}-\frac{\xi_{\textbf{q}}}{2\epsilon_{\textbf{q}}}}=-i\sqrt{1-\frac{J_z}{\epsilon_{\textbf{q}}}},
\end{eqnarray}
where we also defined $\epsilon_{\textbf{q}}=\sqrt{\xi^2_{\textbf{q}}+\left|\Delta_{\textbf{q}} \right|^2}=\sqrt{\xi^2_{\textbf{q}}+\beta^2_{\textbf{q}}}$.\\
In terms of these operators the diagonalised Hamiltonian takes the form 
\begin{equation}
\label{diagoham}
H=\sum_{\textbf{q}}=\epsilon_{\textbf{q}}\left(b^{\dagger}_{\textbf{q}}b_{\textbf{q}}-\frac{1}{2} \right),
\end{equation}
whose ground state has the BCS form
\begin{equation}
\ket{\Psi_0}=\prod_{\textbf{q}}\left(u_{\textbf{q}}+v_{\textbf{q}}C^{\dagger}_{\textbf{q}} C^{\dagger}_{-\textbf{q}}\right) \ket{0},
\end{equation}
which is annihilated by all the $b_{\textbf{q}}$. From the dispersion relation it is possible to find out the phase diagram structure of the system. 
One can readily check that the following triangular inequalities
\begin{align}\label{Bphase}
\left|J_{x} \right| \leq \left|J_{y} \right| &+ \left|J_{z} \right|, \quad \left|J_y \right| \leq \left|J_x \right| + \left|J_z \right|, \\ &\left|J_z \right| \leq \left|J_x \right| + \left|J_y \right|,\nonumber
\end{align}
if satisfied, determine whether the spectrum is gapless.
In Fig. \ref{fig:phasediagram} we explicitly depict the above triangular condition in the positive octant ($J_x, J_y, J_z \geq 0$). One can easily derive the representation in the other octants, by symmetry. The triangular region in the phase diagram determined by the above conditions will be called the gapless B phase, while the other three equivalent regions will be indicated as gapped A phases. \\

\section{Berry curvature in the vortex-free sector on the plane}
\label{berry}

\begin{figure}[h]
\includegraphics[width=\linewidth]{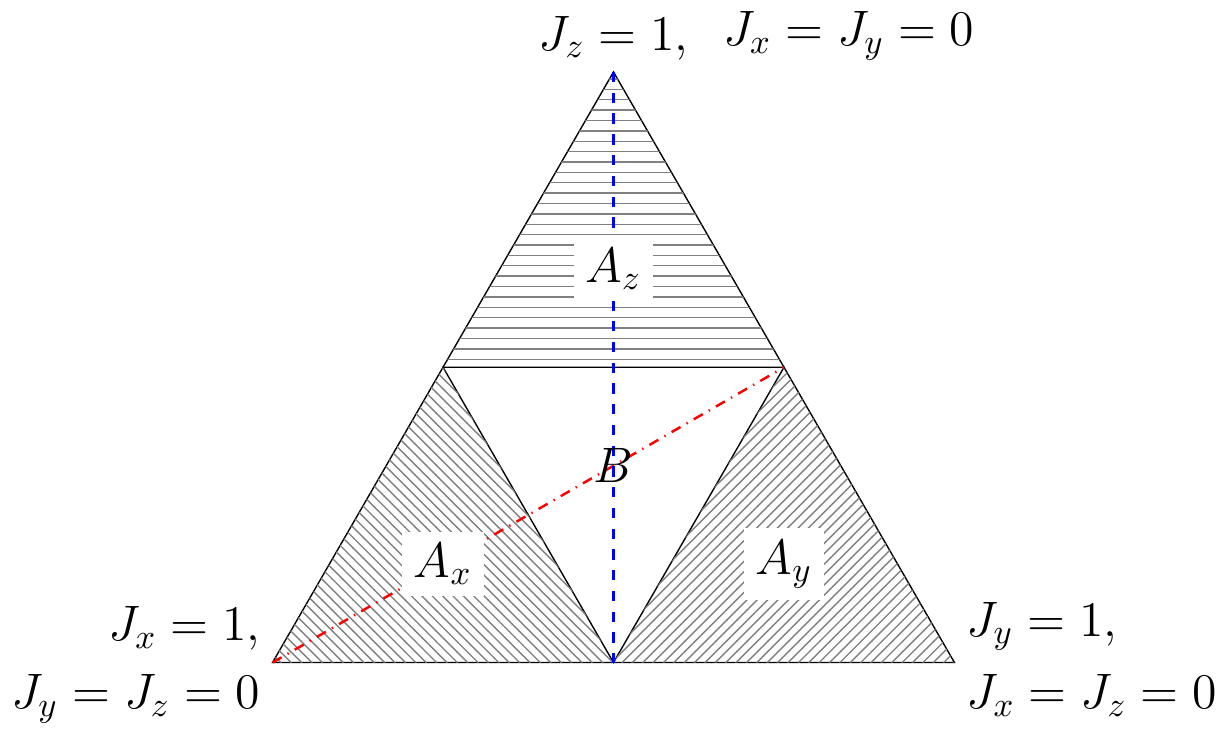}
\caption{Phase diagram: the blue dashed line, taken as the evolution line on which the Berry curvature is explored, is parametrised by $J_x=J_y=J$, while the red dot dashed line is parametrised by $J_y=J_z$.}
\label{fig:phasediagramlines}
\end{figure}

\begin{figure}[h!]
\centering
\includegraphics[width=\linewidth]{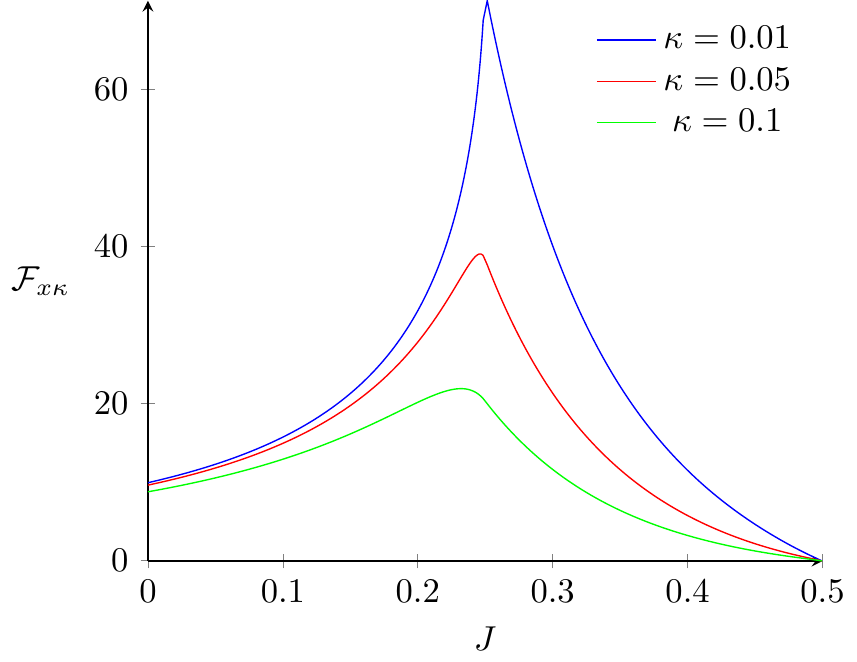}
\caption{$\mathcal{F}_{x \kappa}$ component of the Berry curvature as a function of $J$ along the evolution line $J_x=J_y=J$, $J_z=1-2J$, with external coupling values $\kappa=0.01, \, 0.05, \, 0.1$.}
\label{fig:berry}
\end{figure}

In this section we calculate the Berry curvature $\mathcal{F}_{ij}\left(J \right)= \partial_i \mathcal{A}_j-\partial_j \mathcal{A}_i$, where $\partial_i \equiv \frac{\partial}{\partial J^i}$, of the Berry connection
\begin{equation}
\mathcal{A}_i \left(J \right)=-i\bra{\Psi}\partial_i \ket{\Psi}, \quad i \in \left\{x,y,z \right\}.
\end{equation}
We will focus on the vertex-free configuration in a planar geometry, so we will have to take into account only a single ground state, and therefore we will have to look at an Abelian Berry curvature. \\
Following the Fermionisation procedure used in \cite{Kells2009}, the Hamiltonian in Eq. (\ref{22ham}) can be rewritten explicitly as
\begin{equation}
\label{hampaul}
H_{\textbf{q}}=\textbf{h}(J) \cdot \sigma,
\end{equation}
where $\textbf{h}(J) \equiv \left(0,\, -\beta_{\textbf{q}}, \, \xi_{\textbf{q}} \right)$, and $\sigma$ are the Pauli matrices.
The spectral Berry curvature (at fixed $\textbf{q}$, the total curvature would be obtained summing over all momenta) is easily computed directly by means of the relation
\begin{equation}
\label{curvham}
\mathcal{F}_{ij}=\frac{1}{2h^3}\left[\left(\partial_i \textbf{h}\right) \times \left(\partial_j \textbf{h}\right) \right] \cdot \textbf{h},
\end{equation}
where $h:=\left|\textbf{h} \right|=\epsilon_{\textbf{q}}$ and $\partial_{j}:= \partial/\partial J_{i}$.\\
One can readily check that this curvature appears to be zero everywhere, on account of the time-reversal (TR) and parity (P) symmetries of the model.\\
As discussed in the introduction, adding a TR and/or P symmetry-breaking term in the Hamiltonian in the \emph{gapless} $B$ phase, for instance by means of an external magnetic field, results in a non-vanishing gap opening up. This condition allows for the creation of non-Abelian anyonic excitation. Alternatively, one can add a three-body interaction term (TR and P symmetry breaking ) of the form \cite{Kells2014}
\begin{equation}
\label{hint}
H_{\text{int}}=-\kappa \sum_{\textbf{q}} \sum_{l=1}^4 P_{\textbf{q}}^{(l)},
\end{equation}
where $\kappa$ is the three-body external coupling, and with the second summation running over the four terms 
\begin{equation}
\sum_{l=1}^4 P_{\textbf{q}}^{(l)}=\sigma_1^x \sigma^y_6 \sigma^z_5+\sigma_2^z \sigma_3^y \sigma_4^x+\sigma_1^y\sigma_2^x \sigma_3^z+\sigma_4^y \sigma_5^x \sigma_6^z.
\end{equation}
The Hamiltonian $H_{\textbf{q}}$ in Eq. (\ref{22ham}) remains of the same form, provided a real part is added to $\Delta_{\textbf{q}}$: $\Delta_{\textbf{q}}=\alpha_{\textbf{q}}+i\beta_{\textbf{q}}$, with 
\begin{equation}
\alpha_{\textbf{q}}=4\kappa\left[\sin q_x-\sin q_y \right].
\end{equation}
The diagonalised form of this Hamiltonian is then exactly the same as in Eq.(\ref{diagoham}), but with 
\begin{equation}
\epsilon_{\textbf{q}}=\sqrt{\xi^2_{\textbf{q}}+\left|\Delta_{\textbf{q}} \right|^2}=\sqrt{\xi^2_{\textbf{q}}+\alpha^2_{\textbf{q}}+\beta^2_{\textbf{q}}}.
\end{equation}
We can still write $H_{\textbf{q}}$ in the form of Eq.(\ref{hampaul}), but with a slightly different vector $\textbf{h}(J)\equiv \left(\alpha_{\textbf{q}},\, -\beta_{\textbf{q}}, \, \xi_{\textbf{q}} \right)$, and calculate again the spectral curvature.
Of course, one should embed the $3$-dimensional parameter manifold onto a $4$-dimensional one to include the extra parameter $\kappa$.\\
We find that the only non-vanishing components of the curvature in~Eq.(\ref{curvham}) are the $\mathcal{F}_{i \kappa}=-\mathcal{F}_{\kappa i}$, $i \in \{x,y,z \}$, which are explicitly given by
\begin{align*}
{} & \mathcal{F}_{x\kappa, \textbf{q}}=\frac{\left[\sin q_x -\sin q_y \right]}{2\epsilon^3_{\textbf{q}}}\left[\xi_{\textbf{q}}\sin q_x -\beta_{\textbf{q}}\cos q_x \right], \\ & 
\mathcal{F}_{y\kappa, \textbf{q}}=\frac{\left[\sin q_x -\sin q_y \right]}{2\epsilon^3_{\textbf{q}}}\left[\xi_{\textbf{q}}\sin q_y -\beta_{\textbf{q}}\cos q_y \right], \\ & 
\mathcal{F}_{z\kappa, \textbf{q}}=-\frac{\left[\sin q_x -\sin q_y \right]}{2\epsilon^3_{\textbf{q}}}\beta_{\textbf{q}}.
\end{align*}
In order to obtain the total curvature, the spectral curvature $\mathcal{F}_{i \kappa}$ needs to be summed over all quasi-momenta $\textbf{q}$ (or, in the thermodynamic limit, integrating over $d\textbf{q}$).\\
Without loss of generality, let's choose the octant with $J_i\geq0 \, \, \, \, \forall i \in \{x,y,z\}$. One sees that the three gapped phases $A_i$ are obtained for $J_i > J_j+J_k$, so that, for example, the region $A_x$ is determined by the condition $J_x > J_y+J_z$.  The $B$ phase is instead realised by the conditions~(\ref{Bphase}). The four phases are separated by quantum phase transition lines on which one of the $J_i$ is equal to the sum of the other two (see Fig. \ref{fig:phasediagram}). A TR-P breaking perturbation (for instance the term in Eq. (\ref{hint}) with $\kappa \neq 0$) ) opens up a gap in the otherwise gappless phase $B$. This would make both the $A$s and the $B$ phases gapped, however, a distintive property of the latter, compared to former, is that the $A$ phases host Abelian excitations, whereas the low energy excitation of the $B$ phase satisfy non-Abelian anyonic statistics. Notice that, in the chosen octant, the two phases are separated by the plane $J_x+J_y+J_z=1$, and independently of the phase we are in, the couplings have to satisfy such a normalisation condition. To explore the behaviour of the Berry curvature in the different phases and in particular on the transition lines between them, we can choose to study, without loss of generality, the system along the $J_x=J_y$ line, which basically cuts vertically the triangle diagram (blue dashed line in Fig. \ref{fig:phasediagramlines}).
With this choice of line cut we can explore the dependence of the curvature in the $A_z$ and $B$ phases on $J_z$, with a special focus on the critical line at $J_z=\frac{1}{2}$.  Due the symmetry of the model, cutting along this line accounts for the qualitative behaviour of the whole phase space. Under these conditions we can use $J_x=J_y=J$ and, because of the normalisation relation $J_z=1-2J$, the curvature components are just expressed as functions of $0 \leq J \leq \frac{1}{2}$ along this line (the transition at $J_z=\frac{1}{2}$ is then realised at $J=\frac{1}{4}$).\\
After these substitutions the terms appearing in the expressions for the curvature components can be simplified as follows
\begin{align}
\alpha_{\textbf{q}}=&4\kappa\left[\sin q_x-\sin q_y \right] \nonumber\\
\beta_{\textbf{q}}= &2J\left(\sin q_x +\sin q_y \right), \nonumber\\
\xi_{\textbf{q}}= &2J\left(\cos q_x +\cos q_y\right)+2-4J, \label{eq:alpha-beta}\\
\epsilon_{\textbf{q}}=&\sqrt{\xi^2_{\textbf{q}}+\alpha^2_{\textbf{q}}+\beta^2_{\textbf{q}}}=\nonumber\\ 
=&\left\{8J^2[\cos(q_x-q_y )+1]+16\kappa^2[\sin  q_x- \sin q_y ]^2+\right.\nonumber\\
&\left.+(2-4J)[2+4J(\cos q_x +\cos q_y -1)\right\}^{1/2}\nonumber,
\end{align}
so that the Berry curvature components in the thermodynamic limit get simplified as follows
\begin{equation*}
\mathcal{F}_{i \kappa}(J)=\int_{-\pi}^{\pi}\int_{-\pi}^{\pi} dq_x dq_y \mathcal{F}_{i \kappa, \textbf{q}}(J), 
\end{equation*}
with $i \in {x,y,z}$. Explicitly,
\begin{align}
\mathcal{F}_{x \kappa, \textbf{q}}=& 8 \left(\sin q_x-\sin q_y \right) \times\nonumber\\&\times\left[J \sin(q_x-q_y)+\left(1-2J \right)\sin q_x \right]\epsilon_{\textbf{q}}^{-3}, \nonumber\\  
\mathcal{F}_{y \kappa, \textbf{q}}=& 8\left(\sin q_x-\sin q_y \right) \times\label{comp}\\&\times\left[J \sin(q_y-q_x)+\left(1-2J \right)\sin q_y \right]\epsilon_{\textbf{q}}^{-3},\nonumber\\
\mathcal{F}_{z \kappa, \textbf{q}}=& 8J\left(\sin^2  q_y-\sin^2 q_x \right) \epsilon_{\textbf{q}}^{-3}.\nonumber
\end{align}
However, only one of the above expressions is independent.
Indeed, $\mathcal{F}_{x \kappa}(J)=-\mathcal{F}_{y \kappa}(J)$, as can be seen exchanging the dummy integration variables $q_x \rightarrow q_y$ under the integral, while we can see that $\mathcal{F}_{z \kappa}(J)=0$ by using the same argument. We can therefore limit our analysis to the $\mathcal{F}_{x \kappa}(J)$ component. This is an effect of the specific symmetry of the chosen cut-line. Anyway, had we considered another line, we would have got similar results, but on a different set of components. For instance, if we cut the phase diagram from $A_x$ to the right angle of the $B$ phase (red dot dashed line in Fig. \ref{fig:phasediagramlines}), we get $\mathcal{F}_{y \kappa}(J)=-\mathcal{F}_{z \kappa}(J)$, $\mathcal{F}_{x \kappa}(J)=0$, with $J_y=J_z=J$.\\
The numerical result of the integration along the line with $J_x=J_y=J$ for different values of $\kappa \neq 0$ is shown in Fig. (\ref{fig:berry}). It is interesting to note that the function is peaked close to the criticality, at $J=\frac{1}{4}$, while it is regular enough over the whole region $0 \leq J \leq \frac{1}{2}$. However, for $\kappa \neq 0$ it is expected that the eventual criticality could be not evidenced by the Berry curvature, while they are surely caught by the Chern number.
It is also worth noting that the vertical line in the phase diagram (see Fig. \ref{fig:phasediagramlines}) is travelled downward, so that the $A_z$ phase is covered for $0\leq J < \frac{1}{4}$ while the $B$ phase is covered for $\frac{1}{4} < J \leq \frac{1}{2}$. 

\begin{figure}[h!]
\centering
\includegraphics[width=\linewidth]{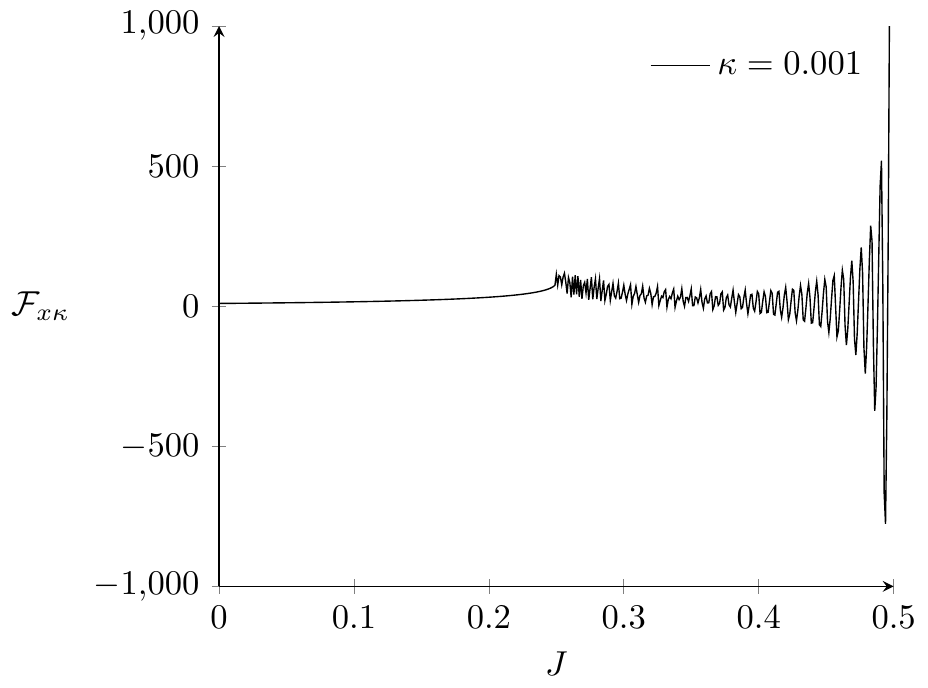}
\caption{$\mathcal{F}_{x \kappa}$ component of the Berry curvature as a function of $J$ along the evolution line $J_x=J_y=J$, $J_z=1-2J$, with $\kappa=0.001$ to resemble the $\kappa \rightarrow 0$ case.}
\label{fig:berryk0}
\end{figure}

The Berry curvature peak gets higher as $\kappa$ decreases to zero. This can be explained on account of the inverse dependence of the Berry curvature on the gap, which, in turn, tightens as $\kappa$ decreases.
To analyse the $\kappa \rightarrow 0$ case, we study the Berry curvature numerically for small enough \footnote{just big enough to avoid numerical instabilities} values of $\kappa$, and we also study analytically the behaviour of the curvature close to the transition line in the $\kappa \rightarrow 0$ limit, estimating the integrals around the Dirac points. This approach is justified by the fact that the dominant contribution to the Berry curvature comes from the regions close to the Dirac points.\\
Therefore, the first thing to do is to find the minima of the energy spectrum around which the integrand function in Eq. (\ref{comp}) can be expanded (we consider again only the $\mathcal{F}_{xk}$ component).
From the analysis of the the function $\epsilon_{\textbf{q}}$ it follows that the two minima are found for the following values of the momentum components
\begin{equation}
q^*_x=-q_y^*=\pm \arccos\left(\frac{1-\frac{1}{2J}}{1-\left(\frac{2\kappa}{J} \right)^2} \right).
\end{equation}
By performing a second order expansion of the integrand function $\mathcal{F}_{x \kappa, \textbf{q}}$ around these minima and using the eigenvalues of the Hessian matrix along the minimum eigendirections we are left to compute the following integral:
\begin{equation}
\int_{-R}^R  \int_{-R}^R dx \, dy \frac{N_0+N_1x^2+N_2y^2}{\left(A^2+B^2 x^2+C^2 y^2\right)^{3/2}}=I_0+I_1+I_2,
\end{equation}
with
\begin{equation}
\begin{aligned}
{} & N_0=-\frac{8}{J^2}\left(J-\frac{1}{4} \right)\left(1-2J \right)\left(\frac{2\kappa}{J} \right)^2, \\ &
N_1=\frac{-40}{J^2}\left(\frac{1}{2}-J \right)\left(J-\frac{1}{4} \right), \\ &
N_2=\frac{8}{J^2}\left(\frac{1}{2}-J \right)\left(J-\frac{1}{4} \right),
\\ &
A=\frac{8\kappa}{J}\sqrt{J-\frac{1}{4}}, \\ &
B=4\left(\frac{1}{2}-J \right), \\ &
C=4\sqrt{J-\frac{1}{4} }.
\end{aligned}
\end{equation}
We also used the fact that the cross terms in the expansion are odd and they do not contribute in the symmetric integration region. The integration variables $x$ and $y$ are the eigencoordinates, i.e. the momentum variables in the basis where the Hessian is diagonal. The finite integration radius $R$ is taken to enclose the minima and its explicit value is not important for the estimate.
It is not hard to see that the contribution coming from $I_0=\int_{-R}^R  \int_{-R}^R dx \, dy \frac{N_0}{\left(A^2+B^2 x^2+C^2 y^2\right)^{3/2}}$ vanishes in the $\frac{\kappa}{J} \rightarrow 0$ limit, while for the other two contributions we find, in the same limit,
\begin{align*}
&\mathcal{F}_x=\lim_{\frac{\kappa}{J} \to 0} \left(I_1+I_2 \right) \\ &\propto\frac{1}{J^2}\left[\frac{\log\left(z+\sqrt{1+z^2} \right)}{z}-5z^2\log\left(\frac{1}{z}+\sqrt{1+\frac{1}{z^2}} \right) \right],
\end{align*}
with $z=\frac{\sqrt{J-\frac{1}{4}}}{\frac{1}{2}-J}$.
The first thing to notice is that in the $J \rightarrow \frac{1}{4}$ limit the Berry curvature is finite, which is in agreement with the numerical analysis.\\
However, even if there is no criticality, the Berry curvatures still gives information about the different phases of the system. In fact, it can be seen numerically that for very small values of $\kappa$ resembling the $\kappa \rightarrow 0$ limit, we find very different behaviours below and above the transition line $J=\frac{1}{4}$. Namely, rapid oscillations appear in the non-trivial phase, as it is showed in Fig. (\ref{fig:berryk0}), explicitly revealing the two different topological phases.
Since the Berry curvature does not show any criticality, it is relevant to analyse also the first derivative of it (w.r.t. the parameter $J$).
With a similar analysis we can estimate the derivative of the curvature, obtaining the following result:
\begin{equation}
\partial_J \mathcal{F}_{x\kappa} \propto \frac{\log\left(J-\frac{1}{4} \right)}{J^2},
\end{equation}
which instead diverges to $-\infty$ in the $J \rightarrow \frac{1}{4}^+$ limit, showing a criticality. The analytical behaviour is in agreement with the numerical result
, which however does not seem to be able to reveal the divergence in the transition point.\\
Therefore, the analysis of the Berry curvature at $\kappa \rightarrow 0$ shows a critical behaviour, revealing the topological phase transition. This was not possible without expanding the parameter space.

\section{Mean Uhlmann curvature estimation at finite temperature}
\label{meanuhlmann}

In this section we get a generalisation of the Berry curvature to a finite temperature case and therefore to a mixed state representation.\\
It was recently introduced in \cite{Carollo2018} a proper mixed state generalisation of the Berry curvature, called mean Uhlmann curvature (MUC) which is gauge-independent and which seems to properly describe topological phase transitions at finite temperature \cite{Carollo2018}, \cite{Leonforte2019}.
It was shown in \cite{Leonforte2019} that for a two-level system described by a Hamiltonian of the form (\ref{hampaul}) in a finite temperature equilibrium state described by the density operator $\rho=\frac{e^{-\beta H}}{Z}$, the MUC can be written as follows
\begin{equation}
\label{muc}
\mathcal{U}_{i j}=\frac{\tanh^3\left(\beta h  \right)}{2h^3}\left[\left(\partial_i \textbf{h}(J) \right) \times \left(\partial_j \textbf{h}(J) \right) \right] \cdot \textbf{h}(J),
\end{equation}
where $\beta=\frac{1}{kT}$ and $Z$ is the partition function.\\
It is clear from Eq. (\ref{muc}) that in this case the MUC is basically the Berry curvature as written in Eq. (\ref{curvham}) with a further $\left(\tanh\left(\beta h \right)\right)^3$ factor, which ensures that in the $T \rightarrow 0$ limit it reduces to the old pure state Berry curvature, while in the high temperature limit ($T \rightarrow +\infty$) the MUC vanishes, as it should be.
Indeed, the MUC can be seen as a kind of statistical average of the curvature of the states. Since the ground and excited states contribute with opposite curvature and at high temperatures the two states tends to be equally populated, it is expected to have $\mathcal{U} \rightarrow 0$.\\
In our case we have a BdG type Hamiltonian (\ref{22ham}) and it was proven in \cite{Leonforte2019} that in this particular case the spectral MUC at fixed momentum $\textbf{q}$ is given by a slightly different expression
\begin{equation*}
\mathcal{U}_{i j, \textbf{q}}=\frac{\tanh \frac{\beta h_{\textbf{q}}}{2} \tanh^2 \beta h_{\textbf{q}}}{2h_{\textbf{q}}^3}\left(\partial_i \textbf{h}_{\textbf{q}}  \times \partial_j \textbf{h}_{\textbf{q}}  \right) \cdot \textbf{h}_{\textbf{q}}.
\end{equation*}
The difference is due to a different normalisation condition. That is due to the fact that we are not really dealing with a two-level system, but this aspect was not effective in the Berry curvature expression because it was related to the ground state and only two states were involved.\\

\subsection{Uhlmann number}

It is also possible to define a so-called \textit{Uhlmann number} in analogy with the Chern number:
\begin{equation}
n_{U} := \frac{1}{2\pi}\iint_{BZ} dq_x dq_y \, \mathcal{U}_{q_x, q_y},
\end{equation}
with, in our case,

\begin{equation*}
\mathcal{U}_{q_x, q_y, \textbf{q}}=\frac{\tanh \frac{\beta h_{\textbf{q}}}{2} \tanh^2 \beta h_{\textbf{q}}}{2h_{\textbf{q}}^3}\left(\partial_x \textbf{h}_{\textbf{q}}  \times \partial_y \textbf{h}_{\textbf{q}}  \right) \cdot \textbf{h}_{\textbf{q}},
\end{equation*}
where the derivatives $\partial_{x}=\partial/\partial q_{x}$, $\partial_{y}=\partial/\partial q_{y}$ are with respect to the components of the quasi-momentum.
As discussed in \cite{Leonforte2019}, $n_U$ is only formally analogue to the Chern number $C$ \footnote{Which in our case can be expressed as $C=\frac{1}{2\pi} \iint_{BZ}dq_x dq_y \, \frac{1}{2h_{\textbf{q}}^3}\left[\frac{\partial}{\partial_{q_x}}\textbf{h}_{\textbf{q}}(J) \times \frac{\partial}{\partial_{q_y}}\textbf{h}_{\textbf{q}}(J) \right] \cdot \textbf{h}_{\textbf{q}}(J)$.} since it is not purely topological and it can be non-integer. However, the two numbers are related by the zero temperature limit, as it has to be: $\lim_{T \rightarrow 0} n_U=C$.\\
Along the cut-line specified in section \ref{berry} we have
\begin{widetext}
\begin{align*}
n_{U} {} & = -\frac{J}{4\pi}\iint_{BZ} dq_x dq_y \, \tanh \frac{\beta \epsilon_{\textbf{q}}}{2} \tanh^2 \beta \epsilon_{\textbf{q}} \frac{J \sin\left(q_x-q_y \right)\alpha_{\textbf{q}}+2\kappa \sin\left(q_x+q_y \right)\beta_{\textbf{q}}+4\kappa\cos q_x\cos q_y\xi_{\textbf{q}}}{\epsilon_{\textbf{q}}^3} ,
\end{align*}
where the $BZ$ is a torus (the momentum $\textbf{q}$ is defined modulo the reciprocal lattice), and the $\alpha_{\textbf{q}}(J)$, $\beta_{\textbf{q}}(J)$, $\xi_{\textbf{q}}(J)$, $\epsilon_{\textbf{q}}(J)$ functions are defined in Eq. (\ref{eq:alpha-beta}).\\
The explicit form of $n_U$ is given by
\begin{align}
n_{U} {} & =-\frac{J\kappa}{\pi} \iint_{BZ} dq_x dq_y \, \tanh\left(\frac{\beta \epsilon_{\textbf{q}}}{2} \right)\tanh^2\left(\beta \epsilon_{\textbf{q}} \right) \times \nonumber\\ & \times \frac{J\left[\cos q_x+\cos q_y\right]+\cos q_x\cos q_y\left(1-2J \right) }{\left\{8J^2\left[\cos\left(q_x-q_y \right)+1\right]+16\kappa^2\left[\sin  q_x- \sin  q_y \right]^2+\left(2-4J \right)\left[2+4J\left(\cos q_x+\cos q_y-1 \right) \right]\right\}^{1/2}}.
\end{align}
\end{widetext}
\begin{figure}[t!]
\begin{tabular}{cc}
  \includegraphics[width=0.5\linewidth]{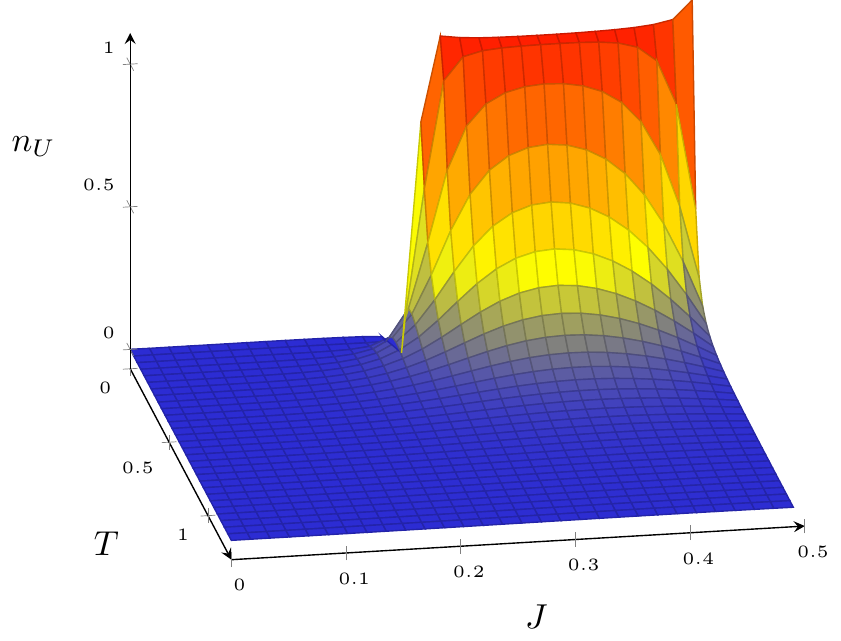}&
  \includegraphics[width=0.5\linewidth]{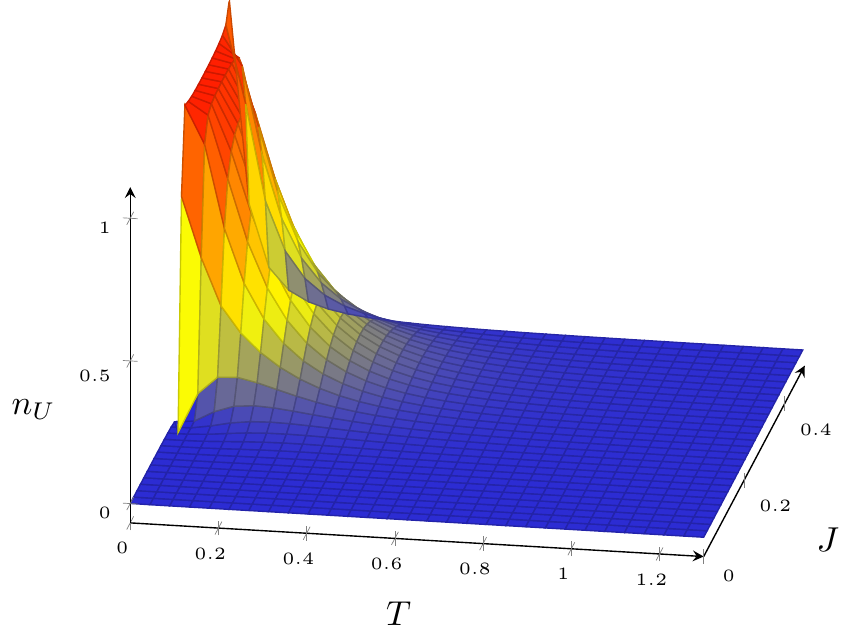}\\
   \includegraphics[width=0.5\linewidth]{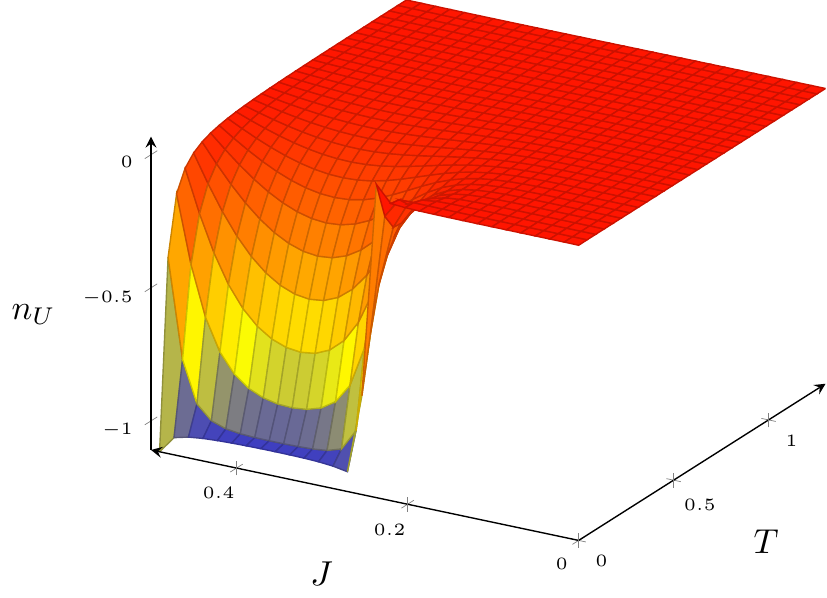}&
  \includegraphics[width=0.5\linewidth]{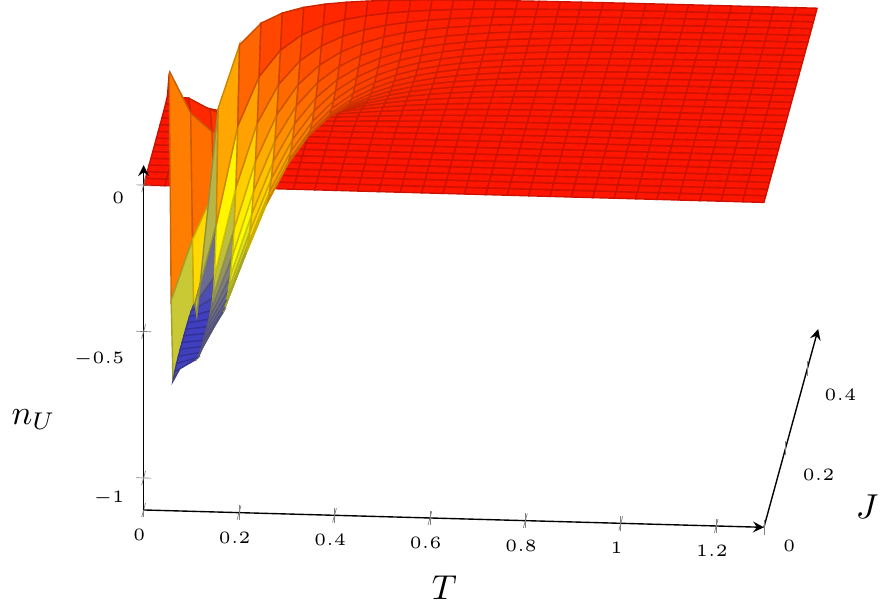}
 \end{tabular} 
\caption{The Uhlmann number $n_{U}$, as a function of $J$ and $T$ (in units of Boltzmann constant, $k_B=1$) along the cut-line $J_x=J_y=J$, $J_z=1-2J$, and two different values of the coupling parameter $\kappa$. First row: front (left) and side (right) view of the Uhlmann number with $\kappa=0.05$. Second row: front (left) and side (right) with $\kappa=-0.05$ }
\label{fig:nu}
\end{figure}
%

In Fig. \ref{fig:nu} is shown the behaviour of the Uhlmann number as a function of the evolution coupling parameter $J$ and temperature $T$, along the cut-line along the cut-line $J_x=J_y=J$, $J_z=1-2J$, for two values of the coupling constant $\kappa$, namely $\kappa=0.05$ and $\kappa=-0.05$.\\
The first thing to notice is the $T \rightarrow 0$ behaviour. The latter reproduces the Kitaev's result for the Chern number calculated by using the projection from the extended Hilbert space~\cite{Kitaev2006}. There, it was found that the Chern number is zero in the $A$ phase, which is topologically trivial, and $\pm 1$ in the $B$ phase. The sign of the Chern number appear to depend on a quantity, which in our case, is the sign of the external magnetic field coupling $\kappa$. Indeed, we find in our case that
%
\begin{equation}
C= \begin{cases}
0, \quad A \, \, \text{phase} \\
1, \quad B \, \, \text{phase}, \, \kappa >0 \\
-1, \quad B \, \, \text{phase}, \, \kappa <0.
\end{cases}
\end{equation}
\begin{figure}[h!]
\centering
\includegraphics[scale=0.9]{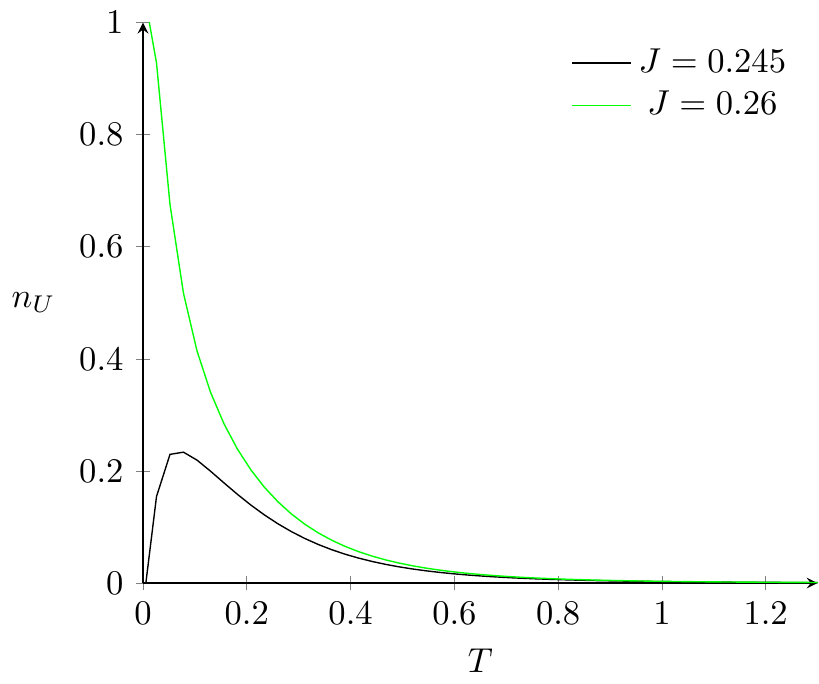}
\caption{Uhlmann number as a function of the temperature slightly below (green) and above (black) the transition point.}
\label{fig:nu2d}
\end{figure}
It is important to note the nonmonotonic behaviour of the Uhlmann number close to $J=\frac{1}{4}$ as a function of the temperature (see Fig. \ref{fig:nu2d}). Specifically, a peak appears for small values of the temperature in the $A$ phase. This effect was also present in the estimation of the Uhlmann number for a p-wave superconductor in \cite{Leonforte2019} and seems to be a natural feature of the Uhlmann number.
To explain this behaviour of the Uhlmann number, we recall that we are working in a two-band system. As a consequence, in the topologically trivial phase ($0 \leq J < \frac{1}{4}$) we have opposite contributions to the curvature, coming from the two bands close to the Dirac points and from the rest of them. Considering the valence band and expanding the Hamiltonian around a Dirac point we find that the region close to this point contributes with a $\pm c$ to the Uhlmann number, while the remaining region gives an opposite contribution $\mp c$. In turn, in the conduction band we have the opposite situation. Indeed, at zero temperature the only contribution comes from the valence band, hence the sum of the two contributions coming from the Dirac points must give $n_{U|_{T=0}}=0$. Increasing the temperature we also get an additional contribution coming from the region of the conducting band close to the Dirac point. Therefore, the situation is not balanced and we have a positive net contribution to the Uhlmann number, that is represented by the peak shown in Fig. \ref{fig:nu2d} (see black curve). At high temperature the main contribution in the valence band comes from the region close to the Dirac point and the same happens for the conduction band. Since their curvature is opposite, the Uhlmann number goes back to zero. This can also be seen considering that, since we are dealing with a two-level system, then $\mathcal{F}_{\text{V.B.}}=-\mathcal{F}_{C.B.}$, and we can  write the Uhlmann curvature as (for sake of simplicity we will suppress every curvature component index and integration measure in the following)
\begin{equation*}
\mathcal{U}=f(\beta\epsilon)\mathcal{F}_g+g(\beta\epsilon)\mathcal{F}_e,
\end{equation*}
where
\begin{align*}
f(x)&=\tanh \left(\frac{x}{2}\right)  \frac{1+\tanh^2 \frac{x}{2}}{2},\\
g(x)&=\tanh \left(\frac{x}{2}\right) \frac{1-\tanh^2\frac{x}{2}}{2},
\end{align*}
so that
\begin{equation*}
n_{U}=\frac{1}{2\pi}\int_{BZ} \mathcal{U}=\frac{1}{2\pi}\left[\int f(\beta \epsilon)\mathcal{F}_g+\int g(\beta \epsilon)\mathcal{F}_e \right],
\end{equation*}
where the $e$ subscript indicates the excited state, while the $g$ subscript stands for the ground state. Then, decomposing this in the contributions coming from the regions of the bands close and far from the Dirac point, a formal description is obtained.
Indeed, decomposing the integration region as $BZ = \Omega_c \cup \Omega_f$, where $\Omega_c$ and $\Omega_f$ are the regions close and far from the Dirac point, we get
\begin{align*}
n_{U}=\frac{1}{2\pi}\left[\int_{\Omega_c} f(\beta \epsilon)\mathcal{F}_g+\int_{\Omega_f} f(\beta \epsilon)\mathcal{F}_g+\right.\\\left.+\int_{\Omega_c} g(\beta \epsilon)\mathcal{F}_e+\int_{\Omega_f} g(\beta \epsilon)\mathcal{F}_e \right].
\end{align*}
In the $\Omega_f$ region, for $T \gtrsim 0$, we can see that $g(\beta \epsilon) \approx 0$, while $f \approx 1$.
Moreover, we can write $f(\beta \epsilon)\mathcal{F}_g=\tanh\left(\frac{\beta \epsilon}{2} \right)\mathcal{F}_g+g(\beta \epsilon)\mathcal{F}_e$, hence
\begin{align}
\nonumber
n_U&=\frac{1}{2\pi}\left[\int_{BZ} \tan\left(\frac{\beta \epsilon}{2} \right)\mathcal{F}_g+\int_{\Omega_c} g(\beta \epsilon)\mathcal{F}_e \right]\\&= C+\frac{1}{2\pi}\int_{\Omega_c} g(\beta \epsilon)\mathcal{F}_e.\label{addit}
\end{align}
In the trivial phase case $C=0$, but for low non-vanishing temperature, an additional positive term, which is responsible for the peak, is present.\\
Finally, the green curve in Fig. \ref{fig:nu2d} describes the Uhlmann number behaviour just outside of the trivial phase, and it simply shows the standard expected behaviour. This behaviour is due to the additional term in Eq. (\ref{addit}), which is negative in this case \footnote{In general, the sign of this term coming from the partial filling of the conduction band is always opposite of that of the Chern number}.

\section{Conclusions}
\label{conclusions}

After reviewing the Kitaev honeycomb model, we mapped the model Hamiltonian to a BdG one and gave explicit relations for the relevant quantities we were interested in. In particular, we assumed a translationally symmetric condition, by considering the vortex-free sector of the model on an infinite plane. In Sec. \ref{berry} we have calculated the Berry curvature by assuming an expanded parameter manifold, which included an extra time-reversal symmetry breaking term, (i.e. an effective magnetic field). This latter perturbation changes the classes of the model from an intrisic topological material to a symmetry protected topological material of class $D$. This was required both at an analytical and conceptual level: on the one hand it allowed for an analytical headway for the calculation of the Berry curvature in the $\kappa \rightarrow 0$ limit, on the other hand it provided a way to properly assign a Chern number to the system, and study the finite temperature case by  the Uhlmann number.
For the $\kappa \rightarrow 0$ case we estimated Berry curvature, by expanding around the relevant Dirac points. We found no criticality from it. However, the first derivative of the Berry curvature shows a divergence in the transition point that signals the phase transition. Therefore, the analysis of the Berry curvature in the $\kappa \rightarrow 0$ limit shows a criticality in the transition line that was not possible to estimate without a parameter expansion.\\
In sec. \ref{meanuhlmann} we calculated the mean Uhlmann curvature, as a generalisation of the Berry curvature at finite temperature, and the Uhlmann number. Indeed, considering a thermal state, the analysis of the Uhlmann number makes it possible to understand how the topology of the honeycomb lattice model evolves as the temperature increases. In particular, no phase transition induced by the temperature is found, but it is shown that the non-trivial phase smoothly disappears at high temperatures. The zero temperature limit correctly reproduces the Chern number result. We also found a nonmonotonic behaviour of the Uhlmann number close to the criticality as a function of the temperature, with a peak appearing for small values of the temperature in the trivial phase. This seems to be a general feature, due to the partial filling of the conduction band. 

This work was supported by the Government of the Russian Federation through Agreement No. 074-02-2018-330 (2), and partially by the Ministry of Education and 
Research of Italian Government.

\vspace{180pt}

\bibliography{ref}
\end{document}